\documentclass[12pt]{JHEP3}
\usepackage{epsfig,amsmath,amssymb}

\newcommand{\lsim}{\mathrel{\mathop{\kern 0pt \rlap
  {\raise.2ex\hbox{$<$}}}
  \lower.9ex\hbox{\kern-.190em $\sim$}}}
\newcommand{\gsim}{\mathrel{\mathop{\kern 0pt \rlap
  {\raise.2ex\hbox{$>$}}}
  \lower.9ex\hbox{\kern-.190em $\sim$}}}

\preprint{KIAS-P09047}

\title{ Dirac gaugino as leptophilic dark matter}

\author{Eung Jin Chun  and Jong-Chul Park \\
\\Korea Institute for Advanced Study, Heogiro 87, Dongdaemun-gu, Seoul 130-722, Korea\\
Emails: \email{ejchun@kias.re.kr, jcpark@kias.re.kr} }
\author{Stefano Scopel \\
\\Department of Physics and Astronomy, Seoul National University \\
Gwanak-ro 599, Gwanak-gu, Seoul 151-742, Korea\\
Email: \email{scopel@phya.snu.ac.kr} }

\abstract{ We investigate the leptophilic properties of Dirac gauginos
  in an R--symmetric N=2 supersymmetric model with extended gauge and
  Higgs sectors. The annihilation of Dirac gauginos to leptons
  requires no chirality flip in the final states so that it is not
  suppressed as in the Majorana case. This implies that it can be
  sizable enough to explain the positron excess observed by the PAMELA
  experiment with moderate or no boost factors. When squark masses are
  heavy, the annihilation of Dirac gauginos to hadrons is controlled by
  their Higgsino fraction and is driven by the $hZ$ and $W^+W^-$ final
  states. Moreover, at variance with the Majorana case, Dirac gauginos
  with a non-vanishing Higgsino fraction can also have a vector
  coupling with the $Z$ gauge boson leading to a sizable
  spin--independent scattering cross section off nuclei. Saturating
  the current antiproton limit, we show that Dirac gauginos can leave a
  signal in direct detection experiments at the level of the
  sensitivity of dark matter searches at present and in the near
  future.}

\maketitle

\begin{document}

\section{Introduction}

N=1 supersymmetry broken around the weak scale would be the prime new
physics candidate to be searched at the LHC. It provides not only an
appealing explanation for the origin of the electroweak symmetry
breaking but also a natural dark matter (DM) particle of the
Universe. While N=1 supersymmetry is enforced by the chiral structure
of matter fields in the Standard Model, the gauge sector may be
extended to have N=2 supersymmetry in which gauginos can be Dirac
particles
\cite{hall91,Randall92,fox02,nelson02,fox04,benakli05,benakli06,
  benakli08,amigo08} (for a review of N=2 supersymmetry, see
\cite{N=2}). Determining the Majorana/Dirac nature of gauginos
will be an interesting task for future experiments looking for
supersymmetric CP and flavor violation~\cite{hisano06,kribs07},
collider signatures \cite{nojiri07,drees08}, and dark matter
properties~\cite{ullio06,hsieh07,kribs08,benakli09}.

There are some important distinctions between Majorana and Dirac
gauginos. First, the annihilation of a Dirac gaugino pair into a
fermion--antifermion pair, $\chi\bar{\chi} \to f\bar{f}$, has a
non-vanishing $s$--wave contribution even in the limit of vanishing
fermion masses, and thus the leptonic final states are not
suppressed. This implies that a Dirac gaugino can provide a viable DM
explanation for the abundance of energetic electrons and positrons
recently observed in cosmic
rays~\cite{HEAT,AMS-01,PAMELA,PPB-BETS,ATIC,HESS,FERMI}. Second, Dirac
gauginos can have a vector coupling with the $Z$ gauge boson,
$\bar{\chi}\gamma_\mu\chi\, Z^\mu$, leading to sizable
spin--independent scattering off nuclei.  As a consequence, the
Higgsino component in the DM gets constrained by the direct detection
data \cite{CDMS} and also by the non-observation of antiproton
excesses in PAMELA \cite{pbar} as will be discussed in detail in the
following.

In order to ensure such Dirac nature of DM, Majorana mass terms,
which provide the mass splitting between the Dirac components,
have to be highly suppressed. Otherwise, the heavier component of
the two quasi-degenerate Majorana gauginos will decay to the
lighter one, and the galactic DM will consist of a pure Majorana
gaugino.  If one assumes the N=2 structure in the Higgs sector,
i.e. that the two Higgses $H_u$ and $H_d$ form an N=2
hypermultiplet \cite{benakli06,benakli08}, the Dirac structure
imposed at the tree level is spoiled by a large amount due to
radiative corrections with Higgs--Higgsino and fermion--sfermion
in the loop.  A way to enforce an (almost) pure Dirac property for
gauginos is then to assume a continuous R symmetry which forbids
the usual $\mu$ term for the Higgs bilinear coupling $H_u H_d$ and
the soft breaking trilinear $A$ term~\cite{kribs07}.

In this paper, we will work out a framework for R--symmetric Dirac
gauginos, which are introduced by extending the gauge/Higgs sector
of a supersymmetric model. We assume the Dirac bino as the main
component of the lightest supersymmetric particle. The
annihilation of two Dirac binos into a lepton--antilepton pair
usually dominates over the annihilation into a quark--antiquark
pair, as sleptons are typically much lighter than squarks. As we
will see, the annihilation to a CP even final or intermediate
state turns out to be velocity--suppressed, which is a property
that is also true in the usual Majorana gaugino case.  Then, the
Dirac bino annihilation to the two $W$ bosons ($W^+W^-$) and that
to a CP even Higgs boson and a $Z$ boson ($hZ$), both controlled
by the Higgsino component in the DM particle, turn out to be the
next important sources of the cosmic antiproton flux.  Quantifying
the suppression factor for the Higgsino component, one can make a
prediction for the direct detection rate.

In Section 2, an R--symmetric N=2 extension of the Minimal
Supersymmetric Standard Model is introduced to enforce an almost pure
Dirac gaugino structure. Based on this framework, we derive
interaction vertices for the Dirac bino, which is assumed to be the
dark matter, and then provide a qualitative analysis for its indirect
and direct detection properties in Section 3. A numerical analysis is
performed in Section 4 to obtain various fits and constraints of the
leptophilic Dirac bino dark matter from the current indirect and
direct detection data. We give our conclusions in Section 5.

\section{The R--symmetric gauge/Higgs sector and Dirac bino couplings}
\label{sec:r_symmetric}

Let us first note that a small Majorana mass term, breaking the
Dirac structure,  will produce quasi-degenerate two Majorana
gauginos,  the heavier of which will then decay to the lighter
one.  Assuming a tiny splitting $\delta m_M$ in the Dirac bino
components, one expects to have one--loop diagrams generating the
magnetic moment operator:
\begin{equation}
    {\alpha' \over 4\pi} { m_{\tilde{B}} \over \tilde{m}^2}
    \tilde{B}_1 \Sigma_{\mu\nu} \tilde{B}_2 F^{\mu\nu}\,,
\end{equation}
between two almost degenerate Majorana components
($m_{\tilde{B}_2} - m_{\tilde{B}_1} = \delta m_M$) where
$m_{\tilde{B}}\approx m_{\tilde{B}_{1,2}}$  is the almost
degenerate bino mass and $\tilde{m}$ is a sfermion mass. The decay
rate for the process $\tilde{B}_2 \to \tilde{B}_1 \gamma$ is
\begin{equation}
 \Gamma \approx {\alpha\over 4\pi} \left( {\alpha'\over 4\pi}
 {m_{\tilde{B}}^2 \over \tilde{m}^2} \right)^2 \delta m_M \,.
\end{equation}
In order to maintain the Dirac gaugino structure for the present
DM, we require that its lifetime is larger than the age of the
Universe. This gives us
\begin{equation} \label{deltamM}
 \delta m_M \lesssim 10^{-33}\, \mbox{GeV}\,,
\end{equation}
for $m_{\tilde{B}} \sim \tilde{m}$.  Thus, the Majorana gaugino
mass term has to be suppressed below the order of $\delta m_M \ll
m_{3/2}^3/M_P^2$, where $m_{3/2}$ is the gravitino mass $\sim 1$
TeV. In N=1 supergravity, a nonzero value of the superpotential
$w_0$ is required to tune the cosmological constant.  As $w_0$
breaks an R symmetry, it can usually generate Majorana gaugino
masses of the order of $m_{3/2} \sim w_0$.  In fact, avoiding such
a leading contribution to gaugino masses and $A$ or $B$  soft
supersymmetry breaking terms has been discussed in the context of
split supersymmetry \cite{arkani04}. Furthermore, a loop
contribution from anomaly mediation was also shown to be further
suppressed \cite{arkani04,luty02}.  Concerning the anomaly
mediation contribution,  it is interesting to remark that,
saturating the lower bound, one finds  $\delta m_M \sim
(g^2/16\pi^2)(1/16\pi^2)(m_{3/2}^3/M_P^2)$ \cite{arkani04} which
can be compatible with the condition (\ref{deltamM}).

An almost R--symmetric supersymmetry breaking can be realized in the
following way. Let us introduce an extended gauge--Higgs sector
\cite{kribs07}:
\begin{equation}
    W'
   = \mu_1 H_d R_u + \mu_2 R_d H_u
   - \sqrt{2} g_a (\xi_1 H_d T^a R_u\,
    + \xi_2 H_u T^a R_d)\, \Phi^a \,,
\label{newgauge}
\end{equation}
where $\Phi^a= (\phi^a, \psi^a)$ is the N=2 counterpart of the gauge
superfield $W^a_\alpha$.  In our phenomenological approach, the N=2
structure is supposed to be broken in an appropriate way leading to an
effective N=1 superpotential, as discussed below.
For instance, $\xi_{1,2}$ in Eq.~(\ref{newgauge}) are arbitrary
parameters breaking the N=2 relation. Note that the usual Higgs
superfields $H_{d,u}$ cannot form an N=2 hypermultiplet as the
corresponding $\mu$ and $B$ terms, breaking an R symmetry, will
generate a large Majorana mass $m_{1/2} \propto (\alpha/4\pi) \mu
B$ at one--loop. A similar one--loop contribution arises also from
the presence of left--right mixing sfermion masses, depending on
the $\mu$ or $A$ term. Thus, we are forced to extend the Higgs
sector to avoid the usual $\mu$ and $B$ terms and we will consider
a situation in which the N=2 structure is not respected in the
extended Higgs sector. Now let us introduce a supersymmetry
breaking field $X$ and two R symmetries, denoted by ${\cal R}$ and
${\cal J}$, which are broken by two order parameters $x_0$ and
$w_0$.  The R charge assignment is given by
\begin{equation}
 \begin{array}{c|ccccccc}
 & W^a_\alpha & \Phi^a & H_{d,u} & R_{u,d} & X & x_0 & w_0 \cr
 \hline
{\cal R} &  1 & 2 & 0 & 0 & 0 & 2 & 2 \cr {\cal J} & 1 & 0 & -1 &
3 & 2 & 0 & 2 \cr
\end{array}\;.
\end{equation}
Note that the two R symmetries are not compatible with the N=2
structure \cite{N=2} as different ${\cal J}$ charges are assigned
to $H_{d,u}$ and $R_{u,d}$ and thus they are not considered to
come from the same hypermultiplets.  The superpotential of the
supersymmetry breaking sector can be written as
\begin{equation}
 W = w_0 + x_0 X\,,
\end{equation}
which gives $\langle X \rangle = \theta^2 F_X$, where $F_X=x_0 \sim
m_{3/2} \sim w_0$ is assumed.  Now one can see that the Dirac
gaugino mass and the extended $\mu$ terms can come from the
following F or D terms:
\begin{equation} \label{eqFD}
 X W^a_\alpha Q_\alpha \Phi^a\big|_F
 \,,\quad H R \Phi\big|_F\;\;\mbox{or}\;\; HRX^\dagger\big|_D\;,
\end{equation}
where $Q_\alpha$ is the supersymmetric covariant derivative. The
last two terms in the above equation reproduce the superpotential
(\ref{newgauge}) and the first term gives the Dirac gaugino mass
term: ${\cal L}_{soft} = M_a \psi^a \tilde{W}^a$. In the
following, we will concentrate on the Dirac gauginos in the
electroweak sector.  The Dirac bino (wino) mass is denoted by
$M_1$ ($M_2$).
The Majorana gaugino mass (and similarly the $A$ term) gets a
contribution from
\begin{equation}
  W^a_\alpha W^a_\alpha W_0^\dagger X X^\dagger\big|_D \,.
\end{equation}
This gives the (Majorana) soft term
${\cal L}_{soft} = \delta m_M \tilde{W}^a \tilde{W}^a$ with
 $\delta m_M \sim  m_{3/2}^3/M_P^2$ which can be
compatible with the bound (\ref{deltamM}) assuming a small
coefficient of the order of $10^{-4}$ for $m_{3/2}\sim 300$ GeV.
Thus, we can safely neglect the Majorana mass terms in our
discussion.

Such a framework allows us to have highly suppressed Majorana mass
terms and results in the following R--symmetric gaugino--Higgsino
mass matrix:
\begin{equation}
\label{eq:mass_matrix}
{\mathcal{M}} \!\! =\!\!
\left[\begin{array}{c
c c c c c}
 0 & M_1 & 0 &  -\xi_1 m_Z s_W c_\beta & 0 & \xi_2 m_Z s_W s_\beta  \\
 M_1 & 0   & -m_Z s_W c_\beta & 0 &  m_Z s_W s_\beta & 0  \\
 0  & -m_Z s_W c_\beta & 0 & \mu_1 & 0 & 0  \\
 -\xi_1 m_Z s_W c_\beta & 0 & \mu_1 & 0 & 0 & 0 \\
  0  &  m_Z s_W s_\beta & 0 & 0 & 0 & \mu_2   \\
 \xi_2 m_Z s_W s_\beta & 0 & 0 & 0 & \mu_2 & 0\\
\end{array}\right]
\end{equation}
in the basis of $(\psi^0, \tilde{B}, \tilde{H}_d, \tilde{R}_u,
\tilde{H}_u, \tilde{R}_d)$. Note that, in order to make a
transparent discussion of the Dirac gaugino properties, we reduced
the number of free parameters by decoupling the wino component
which can be justified in the limit of heavy wino mass: $M_2\gg
M_1,|\mu_i|$.

The Dirac gaugino mass matrix (\ref{eq:mass_matrix}) can be
diagonalized by 6 angles which can be parameterized as
\begin{equation}
{\cal N} =
 \left[
{\begin{array}{cccccc}
c_1  & 0 & s_1 & 0 &  0 & 0 \cr
0 & c_4 & 0 & s_4 & 0  & 0 \cr
-s_1 & 0 & c_1 & 0 & 0 & 0 \cr
0 & -s_4 & 0 & c_4 & 0 & 0 \cr
0 & 0 & 0 & 0 & 1 & 0 \cr
0 & 0 & 0 & 0 & 0 & 1 \cr
\end{array}}
 \right]
 \left[
{\begin{array}{cccccc}
c_2  & 0 & 0 & 0 &  s_2 & 0 \cr
0 & c_3 & 0 & 0 & 0  & s_3 \cr
0 & 0 & 1 & 0 & 0 & 0 \cr
0 & 0 & 0 & 1 & 0 & 0 \cr
-s_2 & 0 & 0 & 0 & c_2 & 0 \cr
0 & -s_3 & 0 & 0 & 0 & c_3 \cr
\end{array}}
 \right]
 \left[
{\begin{array}{cccccc}
1 & 0 & 0 & 0 &  0 & 0 \cr
0 & 1 & 0 & 0 & 0  & 0 \cr
0 & 0 & c_5 & 0 & s_5 & 0 \cr
0 & 0 & 0 & c_6 & 0 & s_6 \cr
0 & 0 & -s_5 & 0 & c_5 & 0 \cr
0 & 0 & 0 & -s_6 & 0 & c_6 \cr
\end{array}}
\right]
.
\end{equation}
The six angles are defined by the diagonalization condition:
${\cal M}_{diag} = {\cal N}^T {\cal M} {\cal N}$,  where the
diagonalized mass matrix ${\cal M}_{diag}$ takes the diagonal form
of $2\times 2$ submatrices which have non-vanishing  symmetric
off-diagonal components of the Dirac mass eigenvalues denoted by
$m_\chi$, $m_{\psi_{H_1}}$, and $m_{\psi_{H_2}}$.

The sizes of the mixing angles play an important role in
determining the cosmic antiproton flux and the direct detection
rate as we will discuss in detail in later sections. It is now
instructive to find their approximate expression in the limit of
small mixing angles: $s_i \ll 1$.  In the leading order of  $m_Z
s_W \ll M_1 < \mu_{1,2}$,  the four mixing angles are given by
\begin{eqnarray} \label{si-approx}
&& s_1 \approx {c_1 m_Z s_W c_\beta (M_1 + \xi_1 \mu_1) \over M_1^2-\mu_1^2} \,,\quad
 s_2 \approx - {c_1 c_2 m_Z s_W s_\beta (c_4^2 M_1 + \xi_2 \mu_2) \over c_1^2 c_4^2 M_1^2- \mu_2^2}\;, \nonumber\\
&& s_3 \approx -{c_3 c_4 m_Z s_W s_\beta (c_1^2 \xi_2 M_1 +\mu_2) \over c_1^2 c_4^2 M_1^2-\mu_2^2} \,,\quad
 s_4  \approx  {c_4 m_Z s_W c_\beta (\xi_1 M_1 +\mu_1) \over M_1^2-
 \mu_1^2}\;,
\end{eqnarray}
and correspondingly the approximate mass eigenvalues are
\begin{eqnarray}
 && m_{\chi} \approx (c_1 c_2 M_1 + s_1 c_2 m_Z s_W c_\beta - s_2 m_Z s_W s_\beta) c_3 c_4 \nonumber\\
&& ~~~~~~  - (c_1 c_2 \xi_2 m_Z s_W s_\beta - s_2 \mu_2) s_3
         +(c_1 c_2 \xi_1 m_Z s_W c_\beta + s_1 c_2 \mu_1) c_3 s_4\;, \nonumber\\
 && m_{\psi_{H_1}} \approx
           (s_1 M_1-c_1 m_Z s_W c_\beta) s_4+(-s_1 \xi_1 m_Z s_W c_\beta+c_1 \mu_1) c_4\;, \nonumber\\
 && m_{\psi_{H_2}} \approx
       (s_3 c_4 M_1+c_3 \xi_2 m_Z s_W s_\beta) c_1 s_2
    +(s_3 c_4 m_Z s_W s_\beta+c_3 \mu_2) c_2\;.
\end{eqnarray}
Also in the limit of small $s_i$ and $\xi_i=1$, the remaining two
mixing angles have the form of
\begin{equation}
s_5 \approx s_6 \approx \frac{m_Z^2 s_W^2 s_\beta}{\mu_1 -
\mu_2}\; \left( \frac{c_\beta}{\mu_1} + \frac{1}{\mu_2} \right)
\;.
\end{equation}

After the diagonalization, one can find the following interaction
vertices of the dark matter particle $\chi$ relevant for our
analysis:
\begin{eqnarray}
\mathcal{L}_{f} &=&
 \sqrt{2}g'  Y_{f_R} {\cal N}_{22} [\overline{f}P_L\chi\widetilde{f}_R  +\overline{\chi}P_Rf\widetilde{f}^*_R] \nonumber\\
 &+& \sqrt{2}g' Y_{f_L} {\cal N}_{22} [\overline{f}P_R\chi^c\widetilde{f}_L
+\overline{\chi}^c P_Lf\widetilde{f}^*_L]\;, \\
\mathcal{L}_{h^0} &=& \frac{g'}{2}  h^0
[ C_1 (\overline{\psi}_{H_1}P_R\chi^c+\overline{\chi}^c P_L \psi_{H_1}) \nonumber\\
&& ~~~~~~~ +  C_2  (\overline{\psi}_{H_2} P_R \chi^c + \overline{\chi}^c P_L \psi_{H_2})
 \nonumber\\
&& ~~~~~~~ + C'_1 (\overline{\psi}_{H_1}P_L\chi^c+\overline{\chi}^c P_R \psi_{H_1}) \nonumber\\
&& ~~~~~~~ +  C'_2  (\overline{\psi}_{H_2} P_L \chi^c + \overline{\chi}^c P_R \psi_{H_2})
\nonumber\\
&& ~~~~~~~ + 2\delta_S \,\overline{\chi}\chi]\;, \label{chi-h} \\
\mathcal{L}_{Z} &=&
 \frac{g}{2c_W}\, Z_\mu
 [ \delta_1' ( \overline{\psi}_{H_1}\gamma_\mu P_L\chi^c
                +\overline{\chi}^c\gamma_\mu P_L \psi_{H_1}) \nonumber\\
&& ~~~~~~ +  \delta_1 ( \overline{\psi}_{H_1}\gamma_\mu P_R\chi^c
                 +\overline{\chi}^c\gamma_\mu P_R \psi_{H_1}) \nonumber\\
&& ~~~~~~ + \delta'_2 ( \overline{\psi}_{H_2}\gamma_\mu P_L\chi^c
                     +\overline{\chi}^c\gamma_\mu P_L \psi_{H_2}) \nonumber\\
&& ~~~~~~ + \delta_2 ( \overline{\psi}_{H_2}\gamma_\mu P_R\chi^c
                            +\overline{\chi}^c\gamma_\mu P_R
                            \psi_{H_2}) \nonumber\\
&& ~~~~~~ + \delta_V^2\, \overline{\chi}\gamma_\mu \chi
          + \delta_{A}^2\, \overline{\chi}\gamma_\mu\gamma_5
          \chi]\;,  \label{chi-Z}
\end{eqnarray}
where
\begin{eqnarray}  \label{couplings}
&& C_1 = s_\alpha{\cal N}_{22} {\cal N}_{33}
         + c_\alpha{\cal N}_{22} {\cal N}_{53} \;,\quad~~~~
C_2= c_\alpha{\cal N}_{22} {\cal N}_{55}
     + s_\alpha{\cal N}_{22} {\cal N}_{35} \;, \nonumber\\
&& C'_1 = \xi_1 s_\alpha{\cal N}_{11} {\cal N}_{44}
         + \xi_2 c_\alpha{\cal N}_{11} {\cal N}_{64}\;,\quad
C'_2= \xi_2 c_\alpha{\cal N}_{11} {\cal N}_{66}
    + \xi_1 s_\alpha{\cal N}_{11} {\cal N}_{46}\;, \nonumber\\
&& \delta_S={1\over2}\; (s_\alpha{\cal N}_{22} {\cal N}_{31} +
c_\alpha{\cal N}_{22} {\cal N}_{51} +\xi_1 s_\alpha{\cal N}_{11}
{\cal N}_{42}+ \xi_2 c_\alpha{\cal N}_{11} {\cal N}_{62} )\;,
\nonumber\\
&& \delta_1 = - {\cal N}_{44} {\cal N}_{42}
              + {\cal N}_{64} {\cal N}_{62}\;,\quad
   \delta_2 = + {\cal N}_{66} {\cal N}_{62}
              - {\cal N}_{46} {\cal N}_{42}\;,  \nonumber\\
&& \delta'_1 = - {\cal N}_{33} {\cal N}_{31}
               + {\cal N}_{53} {\cal N}_{51}\;, \quad
   \delta'_2 = + {\cal N}_{55} {\cal N}_{51}
               - {\cal N}_{35} {\cal N}_{31}\;,  \nonumber\\
&& \delta_V^2 = {1\over2}\; ({\cal N}_{31}^2 - {\cal N}_{51}^2 +
{\cal N}_{42}^2 - {\cal N}_{62}^2)\;,
\nonumber\\
&& \delta_{A}^2 = {1\over2\;} ({\cal N}_{31}^2 - {\cal N}_{51}^2 -
{\cal N}_{42}^2 + {\cal N}_{62}^2)\;.
\end{eqnarray}
Here $Y_{f_{L/R}}$ is the $U(1)$ hypercharge of the
left/right-handed fermion $f_{L/R}$ and the corresponding sfermion
$\tilde{f}_{L/R}$. The angle $\alpha$ is the usual diagonalization
angle of the neutral CP even Higgs bosons. In order to simplify
the discussion in our analysis, we will decouple all heavy Higgs
bosons by assuming $m_A\gg m_Z$, where $m_A$ is the pseudoscalar
Higgs boson mass.

\section{Indirect and direct signals of Dirac neutralino DM}

In the standard Majorana case, the direct annihilation of two
neutralinos to two fermions requires a helicity flip in the final
states and is vanishing in the limit of massless fermions,
$m_f=0$. As a consequence of this, the annihilation cross section
to light fermions is suppressed by a factor of $m^2_f/m^2_\chi$.
On the other hand, when the neutralino is a Dirac particle this
suppression is not present, so that the direct annihilation to
leptons is largely enhanced and can be the dominant annihilation
channel. In this case, the $t$--channel annihilation to quarks
through squark exchange can be suppressed if $m_{\tilde{q}} \gg
m_{\tilde{l}}$~\cite{park09}.

Then, in the presence of a large Higgsino component in the Dirac
neutralino composition, annihilations to gauge and Higgs bosons
can also lead to significant hadronic final states. Among them,
annihilations to quarks through a Higgs particle exchange are
suppressed either by a heavy pseudoscalar Higgs boson in the
decoupling limit, or by a suppressed s--wave contribution in the
non-relativistic limit, an effect which is present also in the
Majorana case.

Under these conditions, we analyze the dominant annihilation
channels and the direct detection bound of the Dirac neutralino
based on the results obtained in the previous section. In
Sections~\ref{subsec:ll_hz} and~\ref{subsec:z}, along with the
dominant annihilation of the Dirac neutralino to leptons we will
discuss the other relevant annihilation channels to hadronic final
states. We will present the neutralino--nucleon cross sections in
Section~\ref{subsec:direct}. The result of this section will be
used in Section~\ref{sec:dirac_neutralino} to constrain the
Higgsino component of the Dirac neutralino.

\subsection{Annihilations to $l\overline{l}$ and $h Z$}
\label{subsec:ll_hz}

First, let us consider the $t$--channel annihilation to leptons
through slepton exchange which has to be the dominant annihilation
channel to explain the observed energetic electron and positron
spectrum. In the case of  $m_{\widetilde{l}_R} \ll
m_{\widetilde{l}_L}$, we have
\begin{equation}
\langle\sigma v\rangle_{l\overline{l}} \simeq \mathcal{N}_{22}^4
\frac{2\pi\alpha^2}{c_W^4}\; Y_{l_R}^4\;
\frac{m_\chi^2}{(m_{\widetilde{l}_R}^2+m_\chi^2)^2}\;.
\label{eq:chichi_ll}
\end{equation}
On the other hand, the case of  $m_{\widetilde{l}_R} \simeq
m_{\widetilde{l}_L} \simeq m_{\widetilde{l}}$ leads to
\begin{equation}
\langle\sigma v\rangle_{l\overline{l}} \simeq \mathcal{N}_{22}^4
\frac{2\pi\alpha^2}{c_W^4}\; Y_l^4\;
\frac{m_\chi^2}{(m_{\widetilde{l}}^2+m_\chi^2)^2}\;,
\label{eq:chichi_ll_same}
\end{equation}
where $Y_l^4 \equiv Y_{l_R}^4+Y_{l_L}^4$. For the three families
of the sleptons having same masses, the above expression must be
multiplied by a factor of 3.

The Dirac gaugino $t$--channel annihilation through Higgsino
exchange produces a Higgs and a $Z$ bosons, which can
significantly contribute to the cosmic antiproton flux. The
annihilation rate for this mode is
\begin{eqnarray} \label{eq:sigmav_hz}
\langle\sigma v\rangle_{h Z} &\simeq&
\frac{\pi\alpha^2}{8c_W^4s_W^2} \;
\frac{\sqrt{m_\chi^2-\overline{m}_Z^2}}{m_\chi^3}  \\
&& \left\{
 \left(\frac{m_\chi^2}{m_Z^2}-1 \right) \left[ (y-z)^2 + (y'-z')^2
-\frac12 \frac{m_Z^2}{m_\chi^2} (z+z')^2 \right]  + \frac32
(y+y')^2  \right\}\;, \nonumber
\end{eqnarray}
where
$$
y = \sum_i \frac{m_\chi m_{\psi_{H_i}} C_i
\delta_i}{m_i^2},\;
y' = \sum_i \frac{m_\chi
m_{\psi_{H_i}} C'_i \delta'_i}{m_i^2},\;
z =
\sum_i \frac{m_\chi^2 C'_i \delta_i}{m_i^2},\;
z' =
\sum_i \frac{m_\chi^2 C_i \delta'_i}{m_i^2}\;,
$$
with $m_i^2 \equiv m_{\psi_{H_i}}^2 + m_\chi^2 -
\overline{m}_Z^2$, $\overline{m}_Z^2 \equiv (m_Z^2+m_h^2)/2$. Here
$m_h^2 - m_Z^2 \ll m_\chi^2$ is assumed.

In Section~\ref{sec:analysis}, we will present numerical results
on the limit of $\langle\sigma v\rangle_{h Z}$ coming from cosmic
antiproton fluxes. It is, however, useful to see a qualitative
behavior of the small mixing limit: $\delta_i, \delta'_i \ll 1$.
Taking $\xi_i=1$ in Eq.~(\ref{si-approx}),  the cross section
(\ref{eq:sigmav_hz}) reduces to a simple approximated form in the
zeroth order of $m_Z^2$:
\begin{equation}
\langle\sigma v\rangle_{h Z} \approx
\frac{\pi\alpha^2}{4c_W^4}\;
\frac{m_\chi^2}{(m_{\widetilde{H}}^2+m_\chi^2)^2}\; \left( 1
-\frac{m_\chi}{m_{\widetilde{H}}} \right)^2 \sin^2(\alpha+\beta)
\;,
\end{equation}
where $m_{\psi_{H_1}} \simeq m_{\psi_{H_2}} \simeq
m_{\widetilde{H}}$ and $M_1 \ll |\mu_i|$ is assumed. The ratio
between (\ref{eq:chichi_ll_same}) and (\ref{eq:sigmav_hz}) becomes
\begin{equation}
{ \langle\sigma v\rangle_{h Z} \over \langle\sigma
v\rangle_{l\overline{l}} } \approx {1 \over 8 Y_l^4}
\frac{(m_{\widetilde{l}}^2+m_\chi^2)^2}{(m_{\widetilde{H}}^2+m_\chi^2)^2}
\left( 1 - \frac{m_\chi}{m_{\widetilde{H}}} \right)^2
\sin^2(\alpha+\beta) \;.
\end{equation}
Assuming now $m_{\widetilde{l}} \simeq m_{\widetilde{H}}$ and
$Y_l^4 = 51/16$ (three even leptophilic), we have:
\begin{equation} \label{cross_ratio}
{ \langle\sigma v\rangle_{h Z} \over \langle\sigma
v\rangle_{l\overline{l}} } < \frac{2}{51} \left( 1 -
\frac{m_\chi}{m_{\widetilde{H}}} \right)^2 < \frac{2}{51} \;.
\end{equation}
Comparing the annihilation cross section to charged leptons
compatible with the PAMELA positron excess~\cite{PAMELA} and the
upper bound on the annihilation cross section to $hZ$ from the
PAMELA antiproton data~\cite{pbar}, presented in
Fig.~\ref{fig:leptophilic} of Section~\ref{sec:analysis}, we
obtain a conservative limit of  $\langle\sigma v\rangle_{h Z}/
\langle\sigma v\rangle_{l\overline{l}} \lsim 0.3$ which can be
easily satisfied, as shown in Eq. (\ref{cross_ratio}). In addition,
even for only one lepton channel, the condition is also satisfied:
$ \langle\sigma v\rangle_{h Z} / \langle\sigma
v\rangle_{l\overline{l}} < 2/17 < 0.3$.

\subsection{Annihilations through Z}
\label{subsec:z}

In the zero--velocity limit, the following annihilation channels
through $s$--channel exchange of a $Z$ boson have non-vanishing
amplitudes. Compared to Eq.~(\ref{eq:sigmav_hz}), these rates
contain an extra factor
 $\sim (m_\chi^2/m_{Z,W}^2)\, \delta^2$ and thus can be  important
in certain parameter regions.

The annihilation rate for $\chi\bar{\chi} \to f\overline{f}$ is
\begin{eqnarray}
\langle\sigma v\rangle_{f\overline{f}} &\simeq&
\frac{2\pi\alpha^2}{c_W^4s_W^4}\; \frac{\sqrt{m_\chi^2-m_f^2}}{m_\chi}\\
&& \left\{\frac{m_f^2}{m_Z^4}g_A^2\delta_A^4 +
\frac{m_\chi^2}{(4m_\chi^2-m_Z^2)^2}\left[2(g_V^2+g_A^2)
+\frac{m_f^2}{m_\chi^2}(g_V^2-2g_A^2) \right] \delta_V^4
\right\}\;, \nonumber
\end{eqnarray}
where $g_V=(T_3-2Q_f s_W^2)/2$ and $g_A=-T_3/2$.
The annihilation rates for $\chi\bar{\chi} \to h Z$ and $W^+ W^-$
are respectively given by
\begin{eqnarray}
\langle\sigma v\rangle_{hZ} &\simeq&
\frac{\pi\alpha^2}{c_W^4s_W^4}\;
\frac{\sqrt{m_\chi^2-\overline{m}_Z^2}}{m_\chi} \label{eq:sigmav_hz_z} \\
&&
\left[\frac{m_\chi^2}{m_Z^4}\left(1-\frac{m_Z^2}{m_\chi^2}\right)\delta_A^4
+\frac{m_\chi^2}{(4m_\chi^2-m_Z^2)^2}\left(1+\frac{2m_Z^2}{m_\chi^2}
\right) \delta_V^4 \right]\;, \nonumber \\
\langle\sigma v\rangle_{WW} &\simeq& \frac{\pi\alpha^2}{s_W^4}\;
\frac{m_\chi \sqrt{m_\chi^2-m_W^2}}{(4m_\chi^2-m_Z^2)^2}
\left(4\frac{m_\chi^4}{m_W^4}+16\frac{m_\chi^2}{m_W^2}-17-3\frac{m_W^2}{m_\chi^2}
\right) \delta_V^4\;. \label{eq:sigmav_ww}
\end{eqnarray}
Finally, there is also an annihilation channel $\chi\bar{\chi} \to
2h$. The annihilation rate for this channel is given by
\begin{eqnarray}
\langle\sigma v\rangle_{2h} \simeq
\frac{\pi\alpha^2}{4c_W^4s_W^4}\;
\frac{(m_\chi^2-m_h^2)^{3/2}}{m_\chi(4m_\chi^2-m_Z^2)^2}\;
\delta_V^4 \;.
\end{eqnarray}
Among these four channels, the process $\chi\bar{\chi} \to 2h$ is
the smallest one. The $hZ$ channel is larger than the
$f\overline{f}$ channel even for the top quark. Thus, the dominant
channels are $hZ$ or $W^+W^-$. The ratio between the two channels
is approximately $\delta_A^4 : \delta_V^4/4$. The ratio between
the cross sections for annihilations to $l\overline{l}$ and
$W^+W^-\; (hZ)$ depends on two parameters, $m_\chi$ and
$\delta_V\; (\delta_A)$. The upper bound on the ratio between the
annihilation rates for $\chi\bar{\chi} \to l\overline{l}$ and $W^+
W^-$ ($hZ$) is approximately 0.5 (0.3), as can be seen from
Fig.~\ref{fig:leptophilic}. Then, comparing (\ref{eq:sigmav_ww})
and (\ref{eq:sigmav_hz_z}) with (\ref{eq:chichi_ll_same}), one
finds the limits: $\delta_V \lesssim 0.3-0.03$ and $\delta_A
\lesssim 0.2-0.02$ depending on $m_\chi$ in the mass range 200 GeV
$\le m_\chi \le$ 2000 GeV. Note that the upper limit of $\delta_V$
is close to the current sensitivity of the direct detection of DM
as will be discussed in the following subsection (see
Fig.~\ref{DirectDetec}).  This will be also confirmed by our
numerical analysis performed in a more general parameter space.

\subsection{Direct detection through $Z$ and $h$ exchange}
\label{subsec:direct}

The vector interaction via $t$--channel $Z$ boson exchange
(\ref{chi-Z}) leads to a {\em spin--independent}
neutralino--nucleon cross section. The cross section for the
neutralino--nucleon vector interaction is given by
\begin{align}
\sigma^{\chi-n, p}_{\rm vector} \simeq \delta_V^4\; \frac{\pi
\alpha^2}{64 s_W^4 c_W^4}\; \frac{\mu_n^2}{m_Z^4}\;
\left[\frac{Z}{A} (1-2s_W^2)-\frac12 \right]^2\;,
\label{eq:neutralino_nucleon_z_exchange}
\end{align}
where $\mu_n$ is the reduced mass for $\chi$-nucleon, and $Z$ and
$A$ are the atomic number and the atomic weight of the target
nucleus, respectively.

%
%
\begin{figure}
\begin{center}
\includegraphics[width=0.82\linewidth]{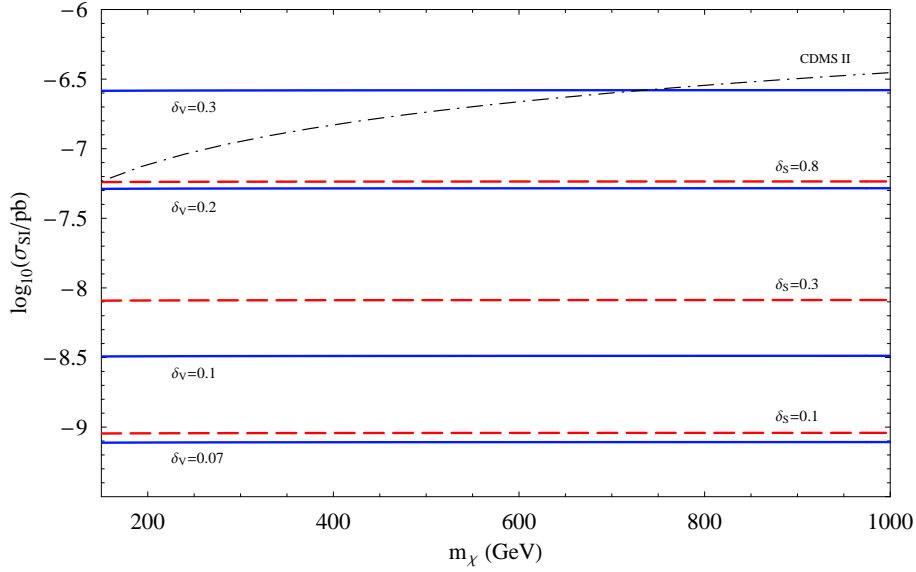}
\end{center}
\caption{Exclusion plot for the spin-independent neutralino--nucleon
cross section $\sigma_{\rm SI}$. The (blue) thick solid lines are
the cross sections via the vector interaction corresponding to
$\delta_V=0.3$, 0.2, 0.1, and 0.07, respectively. The (red) dashed
lines are the cross sections through the Higgs exchange corresponding
to $\delta_S=0.8$, 0.3, and $0.1$, respectively, taking $m_h=115$ GeV.
The dot--dashed line shows the CDMS II limit~\cite{CDMS}.}
\label{DirectDetec}
\end{figure}
%
%

In addition to the vector interaction, the Dirac gaugino $\chi$
interacts with nucleons through $t$--channel Higgs exchange
(\ref{chi-h}). The neutralino--nucleon cross section for the
scalar interaction is
\begin{align}
\sigma^{\chi-n, p}_{\rm scalar} \simeq \delta_S^2\; \frac{0.14^2
\times 4 g'^2 h_{hss}^2}{\pi}\; \frac{\mu_n^2 m_n^2}{m_h^4
m_s^2}\;,\label{eq:neutralino_nucleon_h_exchange}
\end{align}
where $h_{hss}$ is the Higgs--s quark--s quark Yukawa coupling. In
the decoupling limit, this result is estimated as
\begin{align}
\sigma^{\chi-n, p}_{\rm scalar} \simeq \delta_S^2\; \frac{0.14^2
\times 16 \pi \alpha^2}{s_W^2 c_W^2}\; \frac{\mu_n^2 m_n^2}{m_h^4
m_W^2}\;.
\end{align}

In Fig.~\ref{DirectDetec}, we present the neutralino--nucleon
cross sections via the vector interaction (thick solid lines) as a
function of $m_\chi$ for $\delta_V=0.3$, 0.2, 0.1, and 0.07. The
scalar interaction cross sections (dashed lines) are also shown as
a function of $m_\chi$ for $\delta_S=0.8$, 0.3, and 0.1. The limit
from the CDMS II experiment~\cite{CDMS} is shown as a dot--dashed
line.


\section{Analysis}\label{sec:analysis}

In this section, we analyze the phenomenology of Dirac gauginos in
a quantitative way. As discussed in the previous sections, Dirac
gauginos can easily meet the requirements of a leptophilic DM
candidate, and so we will start in Section~\ref{sec:leptophilic}
by summarizing present constraints on generic leptophilic models.
We will then proceed in Section~\ref{sec:bino_leptophilic} to
discuss specific properties of Dirac binos as leptophilic DM,
while in Section~\ref{sec:dirac_neutralino} we will extend the
analysis to Dirac neutralinos of generic composition in some
particular examples.

\subsection{Constraints on leptophilic dark matter}
\label{sec:leptophilic}

\begin{figure}
\begin{center}
\includegraphics[bb=51 230 505 645,width=0.80\linewidth]{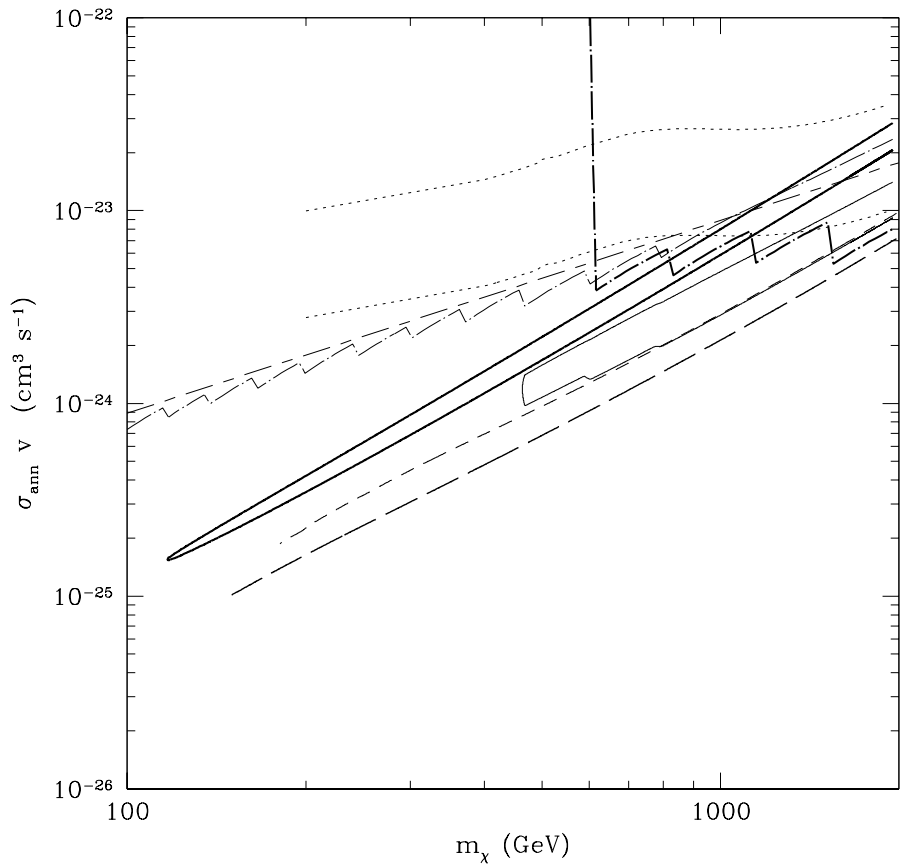}
\end{center}
\caption{Annihilation cross section times velocity $\langle\sigma
  v\rangle$ for a DM candidate annihilating democratically to charged
  leptons of the three families as a function of the mass
  $m_{\chi}$. The thick solid contour shows the range of values
  compatible with the PAMELA positron excess~\cite{PAMELA}; the thin
  solid contour is the range compatible to the observed $e^{+}+e^{-}$
  flux measured by FERMI--LAT~\cite{FERMI}; the thin dot--dashed line
  is the 2 $\sigma$ upper bound from the observed $e^{+}+e^{-}$ flux
  of FERMI; the thick dot--dashed line is the 2 $\sigma$ upper bound
  from the $e^{+}+e^{-}$ HESS measurement~\cite{HESS}; the short
  dash--long dash line represents the upper bound on $\langle\sigma
  v\rangle$ from CMB~\cite{cmb_galli,cmb_slatyer,cmb_padmanabhan}. The
  lower and upper dotted lines show the upper bound on $\langle\sigma
  v\rangle$ obtained by comparing the expected gamma--ray flux
  produced by Inverse Compton (IC) scattering of the final state
  leptons to the FERMI--LAT measurement of the diffuse gamma ray
  emission with or without subtraction of the expected standard
  background, respectively~\cite{FERMI_diffuse}. The short dashed and
  long dashed lines respectively show the upper bounds on the annihilation cross
  section to the final states $W^+W^-$ and $hZ$ from the PAMELA antiproton
  data~\cite{pbar}. See Section \protect\ref{sec:analysis} for
  details.}\label{fig:leptophilic}
\end{figure}

Leptophilic DM annihilation is usually advocated in order to
explain simultaneously the PAMELA positron excess~\cite{PAMELA}
and the excellent agreement between the observed antiproton
spectrum and the corresponding standard expectation \cite{pbar}.
The present situation of a leptophilic DM candidate annihilating
democratically into charged leptons of the three families is
summarized in Fig.~\ref{fig:leptophilic}, where the annihilation
cross section $\langle\sigma v\rangle$ is plotted as a function of
the DM mass $m_{\chi}$.
%
%
In particular, the thick and thin closed solid contours show the
range of values compatible with the PAMELA positron
excess~\cite{PAMELA} and the FERMI--LAT $e^{+}+e^{-}$
data~\cite{FERMI}, respectively. The contours are calculated by
requiring that the $\chi^2$ per degree of freedom for fits to the
data is less than or equal to 1 (only PAMELA data above 10 GeV
have been included in the fit), and assuming for the secondary
positron and electron backgrounds the conventional GALPROP model
denominated by Model \#0 in Table 1 of \cite{fermi09}. As far as
the propagation model for primary electrons and positrons is
concerned, we have adopted a Navarro--Frenk--White (NFW)
profile~\cite{NFW} with medium ranges for the diffusion
coefficient and for the size of the propagation region,
corresponding to the model NFW--med in Table 2 of \cite{cirelli}.
The two dot--dashed lines in the upper part of
Fig.~\ref{fig:leptophilic} represent conservative 2 $\sigma$ C.L.
upper bounds for $\langle\sigma v\rangle$ obtained from the flux
of $e^{+}+e^{-}$ observed by FERMI \cite{fermi09} (thin line) and
the $e^{+}+e^{-}$ flux measured by HESS \cite{HESS} (thick line).

In the same Figure, we also plot for reference with the short
dash--long dash line the upper bound on $\langle\sigma v\rangle$
obtained by considering the imprint on the Cosmic Microwave Background
Radiation (CMB) from the injection of charged leptons from DM
annihilations at the recombination
epoch~\cite{cmb_galli,cmb_slatyer,cmb_padmanabhan}. In particular, the
plotted line is obtained by taking the WMAP5 constraint on the
quantity $f \langle\sigma v\rangle$ from Figure 4 of
Ref.~\cite{cmb_galli}, where $f$ is defined as the average fraction of
the DM rest energy deposited in the gas at $z\simeq$ 800--1000, and
taking $f\simeq 0.39$ from the analysis of Ref.~\cite{cmb_slatyer}. As
one can see, at present the effect of the CMB constraint on
leptophilic models which can explain the PAMELA excess is limited to
very large masses, $m_{\chi}\gsim$ 1.5 TeV. Future polarization data
from Plank will be able to improve this kind of limits
considerably~\cite{cmb_padmanabhan}.

Finally\footnote{Signals from DM annihilation in the Galactic
Center are very sensitive to the choice of density profile.
Therefore, we do not consider them here.}, the dotted lines show
the upper bounds on $\langle\sigma v\rangle$ obtained by comparing
the expected gamma--ray flux produced by Inverse Compton (IC)
scattering of the final state leptons to the diffuse flux of
gamma--rays measured by FERMI at intermediate Galactic
latitudes~\cite{FERMI_diffuse}. We do not include in our analysis
the preliminary data points very recently released by the
FERMI--LAT Collaboration. Due to the large experimental errors,
they would not change our conclusions~\cite{FERMI_preliminary}.
The upper dotted line has been obtained by directly comparing the
flux expected from leptophilic DM (calculated in the approximation
of Ref.~\cite{cirelli_ic}) to the FERMI--LAT data, while the lower
dotted curve has been obtained by subtracting the estimation of
the standard background from the data (as a conservative
estimation of the latter we have assumed the lower boundary of the
dashed region indicated as ``Total'' in the left--hand panel of
Fig.~1 of Ref.~\cite{FERMI_diffuse}).

\begin{figure}
\begin{center}
\includegraphics[bb=23 200 505 645,width=0.80\linewidth]{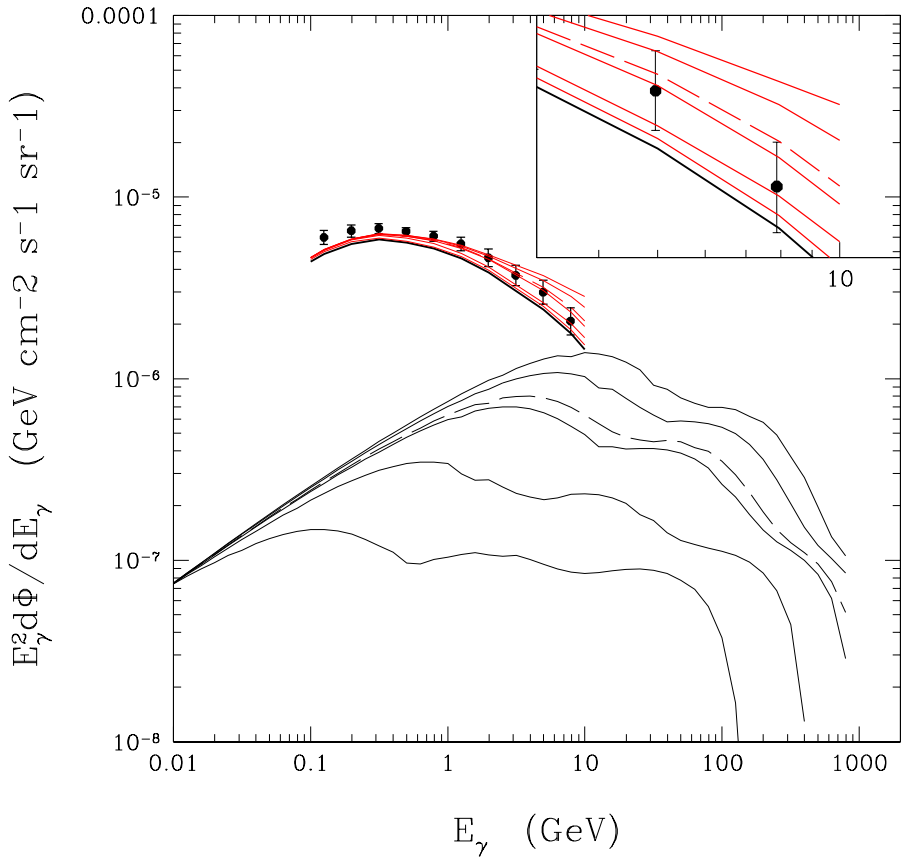}
\end{center}
\caption{ Diffuse gamma--ray flux produced by Inverse Compton
  scattering of the final state electrons produced in DM annihilations as
  a function of the gamma--ray energy $E_{\gamma}$ for a leptophilic
  DM candidate annihilating democratically to charged leptons of the
  three families.  The lower solid curves represent the gamma--ray
  fluxes from DM annihilation for the representative values
  $m_{\chi}$=200, 500, 1000, 1500, and 2000 GeV from bottom to top,
  where for each $m_{\chi}$ the value of $\langle\sigma v\rangle$
  is taken from the lower boundary of the thick solid contour of
  Fig.~\protect\ref{fig:leptophilic} (adopted as a conservative estimation
  for the annihilation cross section required to explain the PAMELA
  data). The same expected fluxes are also summed to the standard
  background, shown by the thick solid line in the range 0.1 GeV $\le
  E_{\gamma} \le$ 10 GeV, and the results are plotted in the same
  energy range. The dashed curves correspond to the IC predictions for
  the additional value $m_{ \chi}\simeq$ 1135 which, as shown in the
  magnified inset at the upper--right, can be adopted as an estimation
  of the upper bound for $m_{\chi}$ from the IC flux from our
  analysis.  The experimental points are the FERMI--LAT data at
  latitudes $10^\circ < b < 20^\circ$ from \cite{FERMI_diffuse}.
}\label{fig:inverse_compton}
\end{figure}

The IC analysis is described in detail in
Fig.~\ref{fig:inverse_compton}, where the diffuse FERMI--LAT data
at latitudes $10^\circ < b < 20^\circ$ from \cite{FERMI_diffuse}
as a function of the gamma--ray energy $E_{\gamma}$ are compared
with the corresponding fluxes from leptophilic DM annihilations.
%
%
In particular, the lower solid curves represent the gamma--ray
fluxes from DM annihilation for the representative values
$m_{\chi}$=200, 500, 1000, 1500, and 2000 GeV (from bottom to
top), where for each $m_{\chi}$ the value of $\langle\sigma
v\rangle$ is taken from the lower boundary of the thick solid
contour of Fig.~\ref{fig:leptophilic} (that we adopt here and in
the following as a conservative estimation for the annihilation
cross section required to explain the PAMELA data). The same
expected fluxes are also summed to the standard background, shown
by the thick solid line in the range 0.1 GeV $\le E_{\gamma} \le$
10 GeV, and the results are plotted in the same energy range. The
dashed curves correspond to the predictions for the additional
value $m_{ \chi}\simeq$ 1135 which, as shown in the magnified
inset at the upper--right, can be adopted as an estimation of the
upper bound for $m_{\chi}$ from the IC flux.

A comment is in order here. Following other similar analysis in
the literature \cite{fermi09,cirelli}, we take the results shown
in Fig.~\ref{fig:leptophilic} as only indicative of the present
trends. In fact, it is clear that a consistent $\chi^2$ analysis
would require a robust assessment of the systematic uncertainties,
an established knowledge of the expected backgrounds\footnote{At
  variance with antiprotons, the expected standard background for
  electrons and positrons could in principle be only in part of
  secondary origin, with a sizable primary fraction (see for
  instance \cite{profumo,blasi}).} and a full marginalization over
the astrophysical parameters, something which is not possible in
light of the present uncertainties.

\subsection{The Dirac bino as a leptophilic DM candidate}
\label{sec:bino_leptophilic}

\begin{figure}
\begin{center}
\includegraphics[bb=51 230 505 645,width=0.70\linewidth]{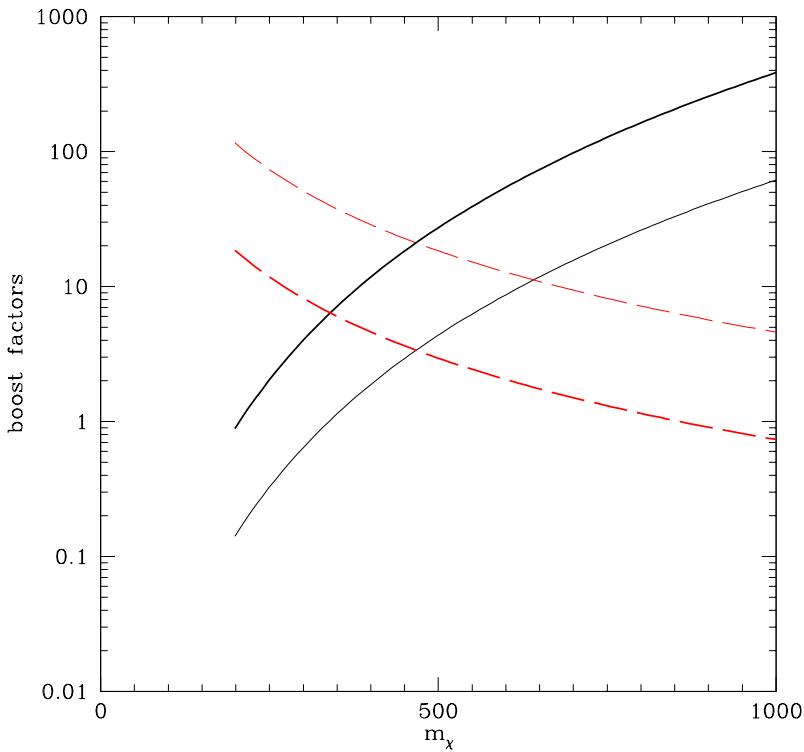}
\end{center}
\caption{Boost factors for a Dirac gaugino DM candidate annihilating
  democratically to leptons of the three families as a function of the
  mass and in the limit $|\mu|\gg M_1$. Solid lines show the ratio
  between the annihilation cross section required to explain the
  PAMELA excess and the value of $\langle\sigma v\rangle$ calculated using
  Eq.~(\protect\ref{eq:chichi_ll_same}). Dashed lines show the ratio
  between the observed lower value of the DM relic density
  $\Omega_{min} h^2=0.098$ \protect\cite{WMAP} and the corresponding
  thermal relic abundance for the Dirac gaugino. Thin lines
  refer to $m_{\widetilde{l}}/m_{\chi}=$ 1 (where we have assumed a small
  mass splitting between $m_{\widetilde{l}}$ and $m_{\chi}$ sufficient to
  avoid coannihilations between sleptons and neutralinos) while thick
  ones to $m_{\widetilde{l}}/m_{\chi}=$2, where $m_{\widetilde{l}}$ is the slepton
  mass.}\label{fig:boost_factor}
\end{figure}

Dirac gauginos can easily meet the requirements of a leptophilic
DM candidate, since their annihilation cross section to leptons
can be dominant, not being chirality flip suppressed as in the
case of Majorana gauginos, while annihilations to quarks
potentially capable of producing an excess in the antiproton
signal can be inhibited by assuming large masses for the squarks
\cite{park09}. If the additional hadronic annihilation channels
discussed in Sections~\ref{subsec:ll_hz} and~\ref{subsec:z} are
further suppressed by assuming a negligible Higgsino component in
the neutralino composition (by taking $|\mu_{1,2}| \gg M_1$ in the
mass matrix of Eq.~(\ref{eq:mass_matrix})), a totally leptophilic
model is obtained for which only the constraint $m_{\chi}\lsim$ 1
TeV from the IC flux discussed in Fig.~\ref{fig:inverse_compton}
is applied\footnote{Notice
  that the IC gamma--ray flux retains the directional information of
  the source, allowing in principle to establish whether the e$^{\pm}$
  PAMELA excess is diffused in all the Galactic halo or localized, for
  instance, in a DM clump close to the Solar System. The preliminary
  data of Ref.~\cite{FERMI_diffuse} do not provide this piece of
  information since they are averaged over all galactic longitudes
  $0^\circ \le l \le 180^\circ$.}. The phenomenology of such a DM
candidate is quite simple, and is summarized in
Fig.~\ref{fig:boost_factor}, where the solid lines show the ratio
between the annihilation cross section $\langle\sigma v\rangle$
required to explain the PAMELA data (taken as before from the lower
boundary of the thick closed solid contour in
Fig.~\ref{fig:leptophilic}) and that calculated by making use of
Eq.~(\ref{eq:chichi_ll_same}) for $m_{\widetilde{l}}/m_{\chi}=$1 (thin
line) and 2 (thick one).
%
%
In particular, $m_{\widetilde{l}}/m_{\chi}\gsim $1 maximizes the
expected annihilation cross section by assuming the lightest
possible slepton mass compatible with the requirement that the
neutralino is the Lightest Supersymmetric Particle (LSP). In this
latter case, the Dirac gaugino annihilation cross section to
leptons can explain the PAMELA data for $m_{\chi}\simeq$ 330 GeV,
while $m_{\chi}\simeq$ 200 GeV is needed if
$m_{\widetilde{l}}/m_{\chi}=$2.

It is well known, however, that the values of the annihilation
cross section that can explain the PAMELA excess ($\langle\sigma
v\rangle \simeq 10^{-24}$ cm$^3$ s$^{-1}$) are about two orders of
magnitude larger than the range compatible to a standard thermal
relic abundance in agreement with observation (2--3$ \times
10^{-26}$ cm$^3$ s$^{-1}$ at the decoupling temperature $T\simeq
m_{\chi}/20$). This is naively shown in
Fig.~\ref{fig:boost_factor}, where the dashed lines represent the
ratio between the 2 $\sigma$ lower value of the observed DM relic
density $\Omega_{min} h^2=0.098$~\cite{WMAP} and the corresponding
thermal relic abundance for the Dirac gaugino. The thin dashed
line refers to $m_{\widetilde{l}}/m_{\chi}=$ 1 while the thick one
to $m_{\widetilde{l}}/m_{\chi}=$ 2. From this figure, one can see
that a light Dirac gaugino ($m_{\chi}\lsim$ 500 GeV) can explain
the PAMELA data with moderate or no boost factor at all if, for
instance, some non-standard expansion history of the Universe
\cite{quintessence} or non-thermal production mechanism is
advocated in order to reconcile its DM relic abundance to the
observation. Alternatively, a thermal relic density in agreement
with observation is attained for $m_{\chi}\gsim$ 800 GeV, but in
this case a large enhancement of the annihilation cross section to
leptons is necessary. Such an enhancement could be provided by an
astrophysical boost factor due to substructures, although recent
analysis seems to disfavor values larger than order 10
\cite{clumpiness}. Furthermore, this piece of information can be
combined to the data on $e^++e^-$ from FERMI--LAT, which, as shown
by the thin closed contour plot in Fig.~\ref{fig:leptophilic},
imply a lower bound on the neutralino mass $m_{\chi} \gsim$ 465
GeV.  Notice, however, that lower values for the mass of the
annihilating DM particle can be in principle assumed if some
additional contribution to the electron+positron background is
claimed at energies $E\gsim 1$ TeV to explain the FERMI--LAT data
independently~\cite{profumo,blasi,extragalactic_source}.

\subsection{The case of a Dirac neutralino of general composition}
\label{sec:dirac_neutralino}

\begin{figure}
\begin{center}
\includegraphics[bb=51 220 505 645,width=0.80\linewidth]{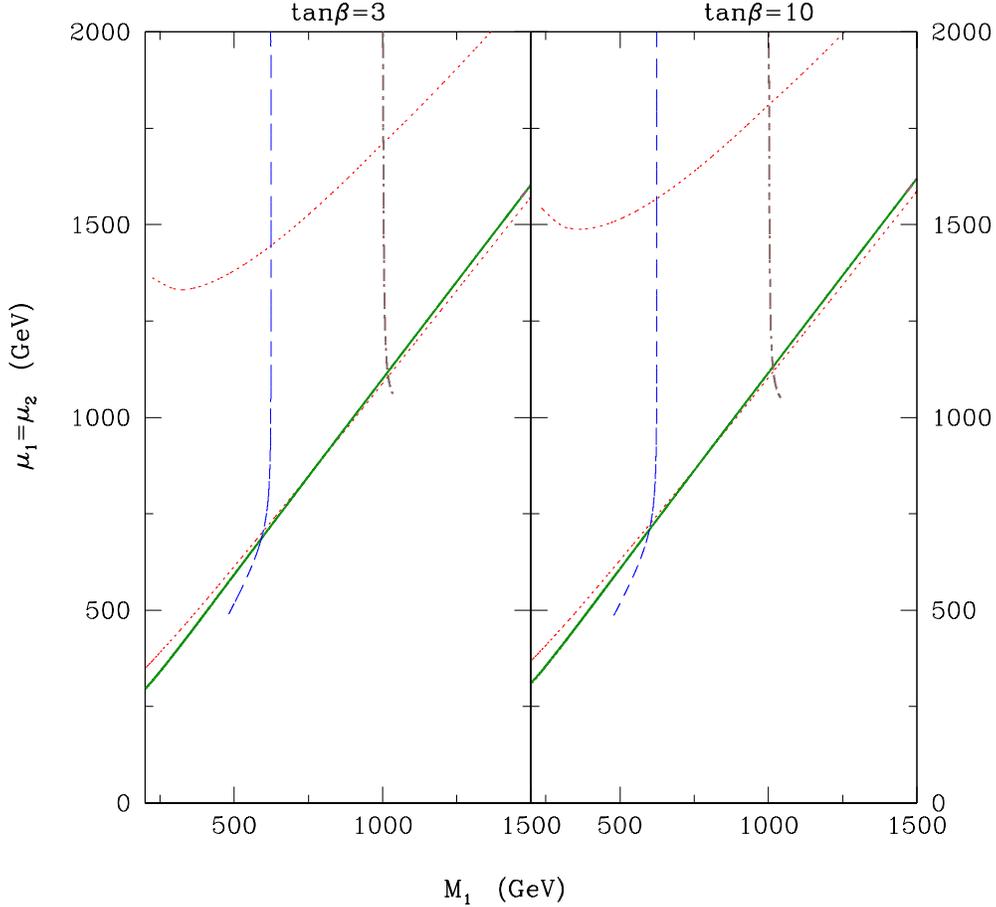}
\end{center}
\caption{Parameter space for a Dirac neutralino of general
  composition, for $\mu_1$=$\mu_2$ and $\tan\beta$=3,10 (left and
  right panels, respectively). In each panel, the region below the
  slanted solid line is excluded by the PAMELA antiproton data, while
  along the two dotted lines the ratio between the neutralino--nucleon
  scattering cross section calculated with
  Eqs. [(\protect\ref{eq:neutralino_nucleon_z_exchange}),
    (\protect\ref{eq:neutralino_nucleon_h_exchange})] and the
  experimental limit from CDMS II~\cite{CDMS} is fixed to 1 (lower
  curve) and 10$^{-3}$ (upper curve). The regions below the lower
  dotted curves, where this ratio is larger than 1, are excluded by
  the direct detection constraint. Moreover, in each plane, the
  vertical thin dashed line to the left represents points where the
  boost factor, defined as the ratio between the annihilation cross
  section $\langle\sigma v\rangle$ required to explain the
  PAMELA excess and that
  calculated by making use of Eq.~(\protect\ref{eq:chichi_ll_same}),
  is equal to 10, while the vertical thick dot--dashed line to the right
  represents points at constant mass $m_{\chi}$=1 TeV.}
\label{fig:m1_mu_ratio_1}
\end{figure}

\begin{figure}
\begin{center}
\includegraphics[bb=51 220 505 645,width=0.80\linewidth]{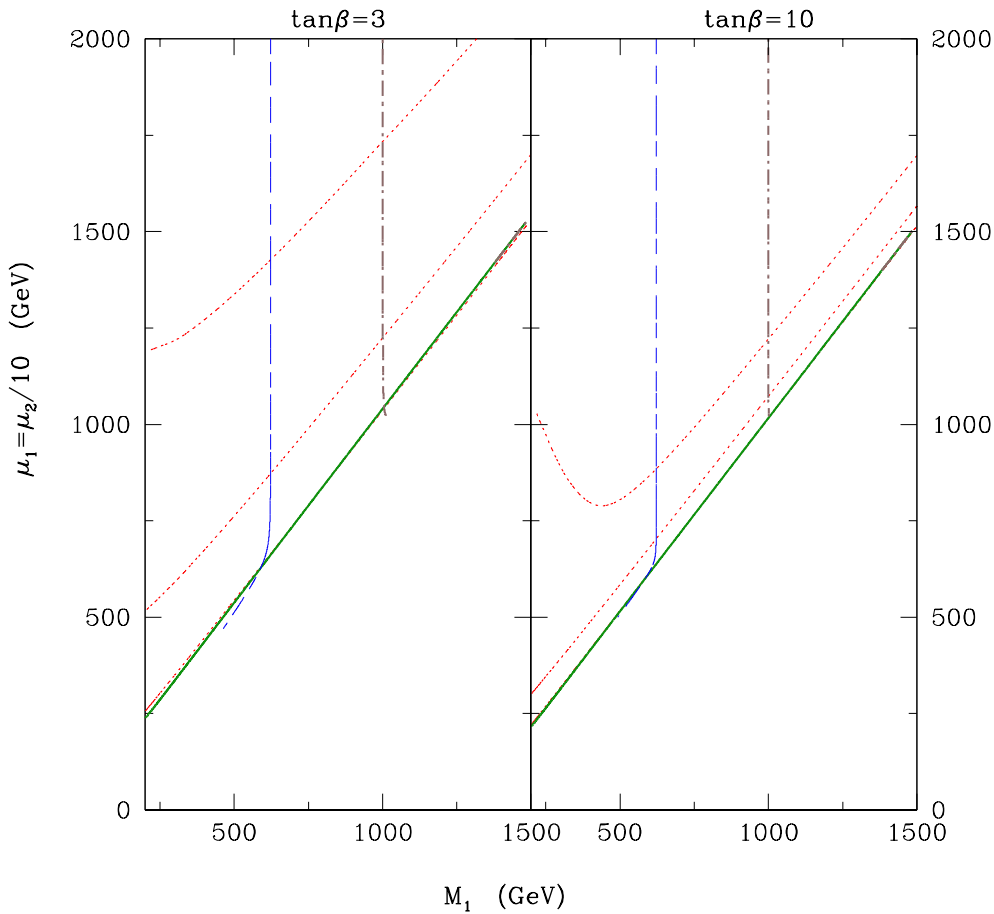}
\end{center}
\caption{The same as in Fig.~\protect\ref{fig:m1_mu_ratio_1}, with
  $\mu_2/\mu_1$=10. In both panels, the three dotted lines show points
  where the ratio between the neutralino--nucleon scattering cross
  section calculated with
  Eqs. [(\protect\ref{eq:neutralino_nucleon_z_exchange}),
    (\protect\ref{eq:neutralino_nucleon_h_exchange})] and the
  experimental limit from CDMS II~\cite{CDMS} is fixed to 1 (lower
  curve, barely visible in the right--hand panel), 10$^{-3}$ (middle
  curve), and 10$^{-5}$ (upper curve).  }\label{fig:m1_mu_ratio_0.1}
\end{figure}

As discussed in Section~\ref{subsec:direct}, an important
phenomenological feature of a Dirac neutralino is the presence of
a vector coupling with the $Z$ boson, which on the other hand is
vanishing in the Majorana case. This implies that, while in the
standard Majorana scenario a diagram with the $Z$ boson exchange
can only contribute to the spin--dependent neutralino--nucleon
cross section,\footnote{The most stringent limits on pure
  WIMP--proton and WIMP--neutron spin--dependent cross sections
  are given by KIMS~\cite{KIMS} and ZEPLIN-III~\cite{ZEPLIN-III_SD},
  respectively.} for the Dirac case it can lead to a much more
sizeable cross section enhanced by a factor of $\simeq A^2$
(see Eq.~(\ref{eq:neutralino_nucleon_z_exchange})). However, note
that the coupling between a neutralino and a $Z$ boson vanishes
(as long as the coupling to the Higgs bosons) in the limit
$|\mu_{1,2}|\gg M_1$ of a pure gaugino. This implies that the
detection of Dirac neutralinos through direct searches is possible
only in the presence of a Higgsino component in a Dirac neutralino
of arbitrary composition.  In order to explore this possibility,
we discuss in this section the full parameter space of our model,
by allowing $|\mu_{1,2}|\simeq M_1$, and discuss the constraints
coming from the hadronic annihilation channels discussed in
Sections \ref{subsec:ll_hz} and \ref{subsec:z}. In particular, in
Figs. \ref{fig:m1_mu_ratio_1} and \ref{fig:m1_mu_ratio_0.1}, the
$M_1-\mu_{i}$ parameter space of the Dirac neutralino is explored
for the two representative cases $\mu_1/\mu_2=1$ and
$\mu_1/\mu_2=0.1$. Our numerical analysis is performed in the case
of $\xi_1=\xi_2=1$.
%
%
%
%
In each figure, the left--hand panel corresponds to $\tan\beta=3$
and the right--hand panel to $\tan\beta=10$. In all the
$M_1$--$\mu_i$ planes, the predominantly Higgsino region below the
slanted solid line is excluded by an antiproton flux above the
PAMELA data of Ref.~\cite{pbar}, driven by the cross section to
$W^+W^-$ of Eq.~(\ref{eq:sigmav_ww}) and a subdominant
contribution from the hZ final state of Eq.~(\ref{eq:sigmav_hz})
(in the calculation of the latter cross section we have assumed a
supersymmetric Higgs sector in the decoupling limit, $m_{H,A}\gg
m_h$, with $m_h$=115 GeV). The corresponding upper bounds to these
two cross sections, which as already mentioned in Sections
\ref{subsec:ll_hz} and \ref{subsec:z} turn out to be the dominant
hadronic ones in our R--symmetric scenario, are plotted for
reference in Fig.~\ref{fig:leptophilic} as short--dashed and
long--dashed lines, respectively. They have been obtained by
conservatively requiring $\chi^2<40$ when comparing the
theoretical expectation to the PAMELA data (this value corresponds
approximately to a 99.5 \% C.L. for 17 degrees of freedom). For
the calculation of the antiproton fluxes, we have adopted the
NFW--med model for propagation \cite{cirelli} and the
parametrization of Ref.~\cite{pbar_background} for the secondary
antiproton background.

In Figs.~\ref{fig:m1_mu_ratio_1} and \ref{fig:m1_mu_ratio_0.1},
the vertical thin dashed line to the left represents points where
the boost factor, defined as the ratio between the annihilation
cross section $\langle\sigma v\rangle$ required to explain the
PAMELA excess and that calculated by making use of
Eq.~(\protect\ref{eq:chichi_ll_same}), is equal to 10. In
particular, the points to the right of this vertical line require
a larger value. Note that an astrophysical enhancement to the
effective cross section provided by DM substructure larger than
order 10 appears to be unlikely \cite{clumpiness}. Moreover, in
each $M_1$--$\mu_i$ plane, the vertical thick dot--dashed line to
the right represents points with a constant Dirac neutralino mass
$m_{\chi}$=1 TeV. Configurations to the right of this line have
larger masses and appear to be disfavored by the IC flux shown in
Fig.~\ref{fig:inverse_compton}. Note that this is the only
irreducible astrophysical bound which is constraining also in the
pure gaugino limit $|\mu_i|\gg M_1$.

In the regions of Figs.~\ref{fig:m1_mu_ratio_1}
and~\ref{fig:m1_mu_ratio_0.1} that are excluded by antiproton
fluxes (below the slanted solid lines), a sizeable Higgsino
fraction can also drive the neutralino--nucleon cross section
above the present experimental limit. This is shown by the dotted
lines which represent configurations with a fixed ratio between
the expected neutralino--nucleus cross section and the present
experimental upper bound from CDMS II~\cite{CDMS}. In particular,
in all the $\mu_i-M_1$ planes this ratio is equal to 1, 10$^{-3}$,
and 10$^{-5}$ (if present), starting from the lower curve. In the
regions below the lower dotted curves, this ratio is larger than
one, so that they are excluded by the present limit from CDMS II.

\begin{figure}
\begin{center}
\includegraphics[bb=51 220 505 645,width=0.4\linewidth]{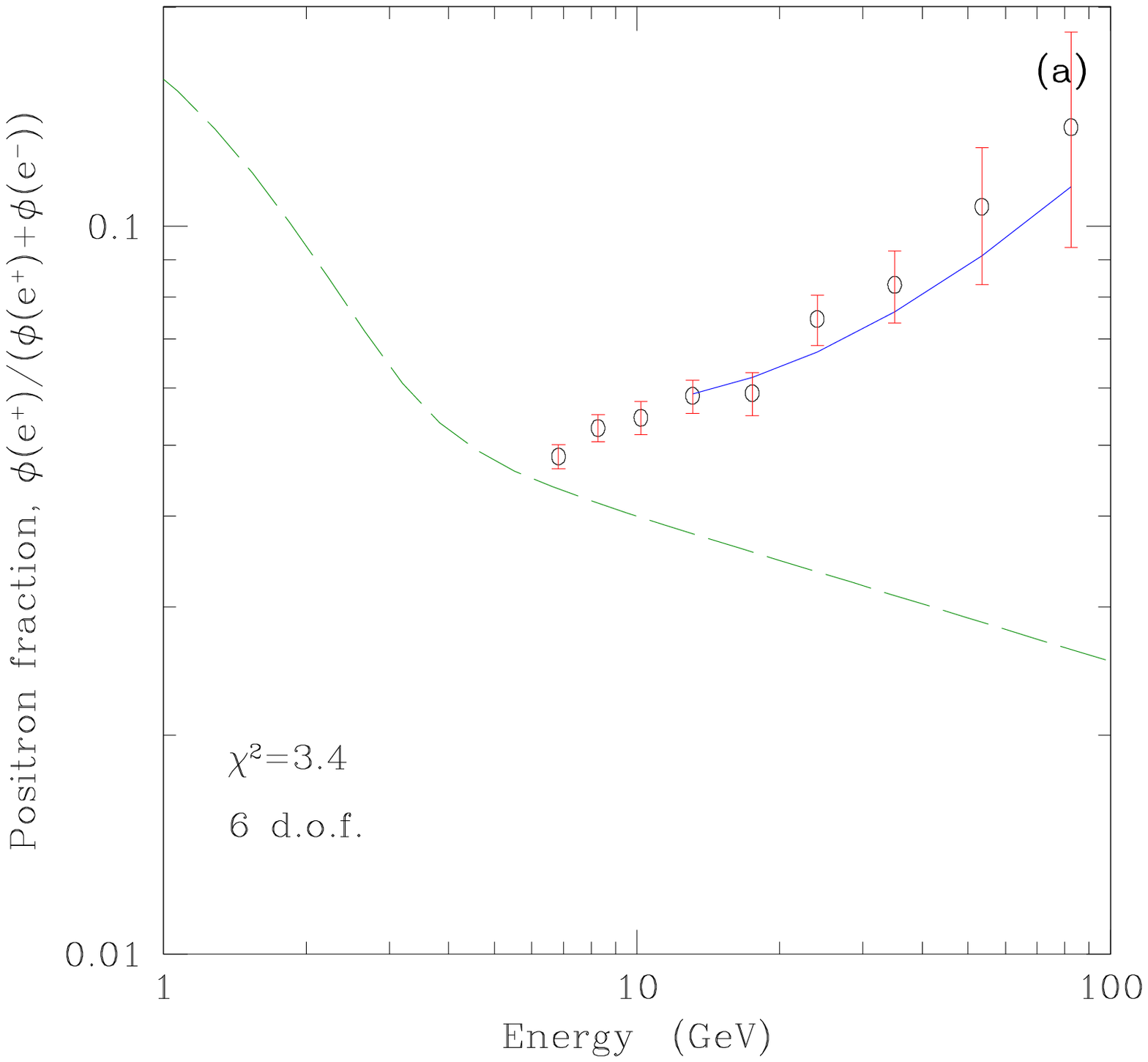}
\includegraphics[bb=51 220 505 645,width=0.4\linewidth]{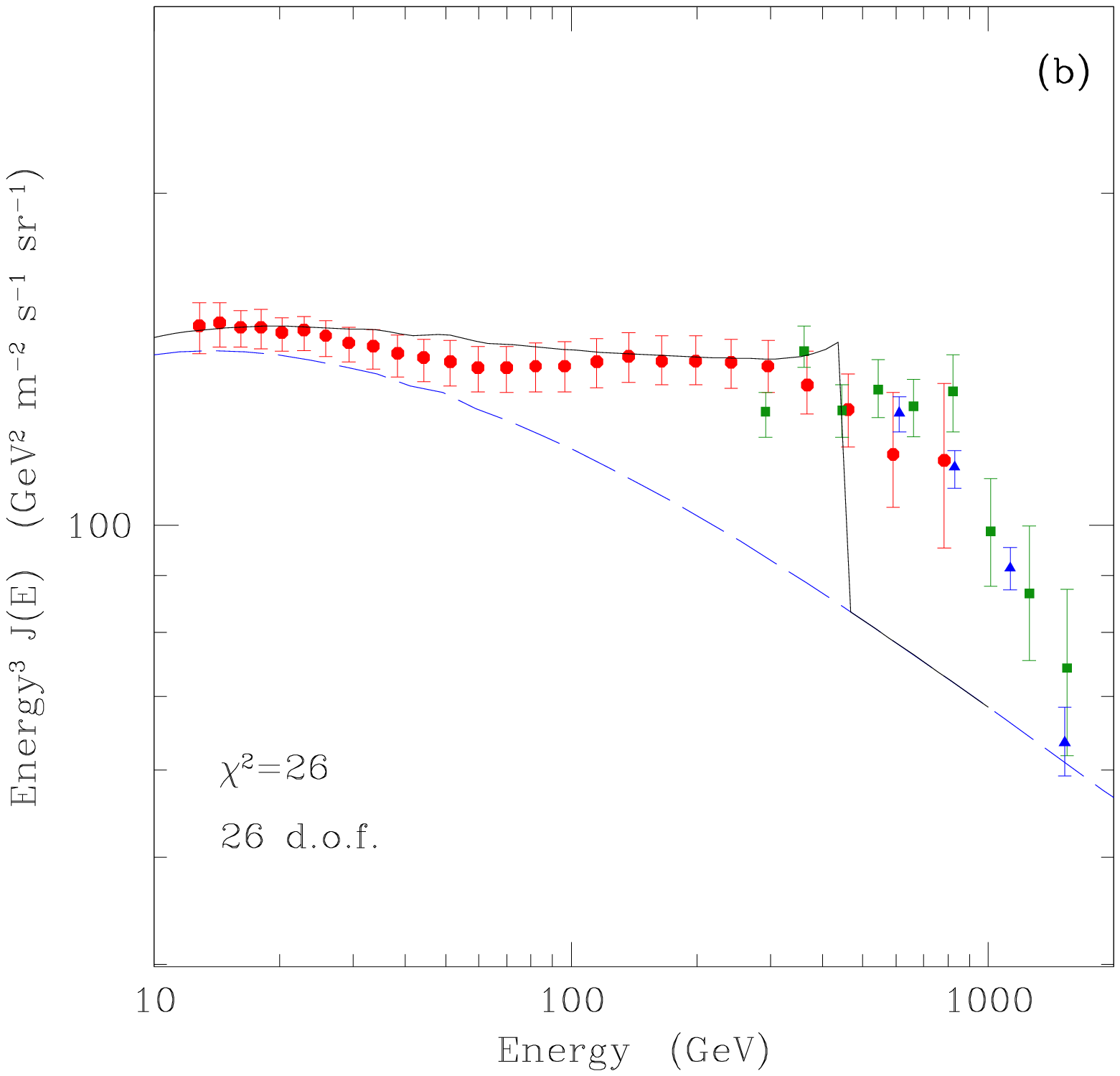}

\includegraphics[bb=51 220 505 680,width=0.4\linewidth]{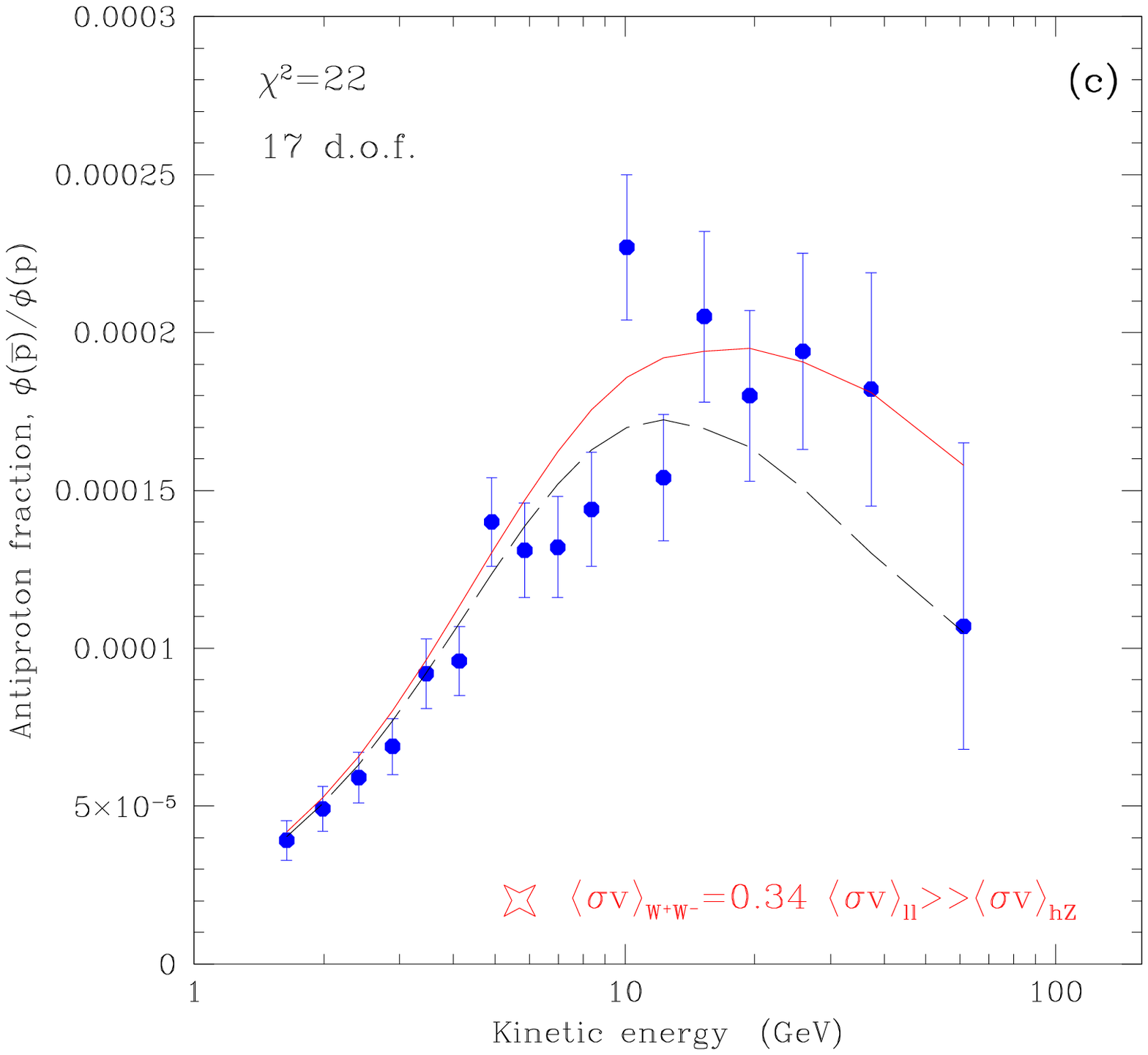}
\includegraphics[bb=51 220 505 680,width=0.4\linewidth]{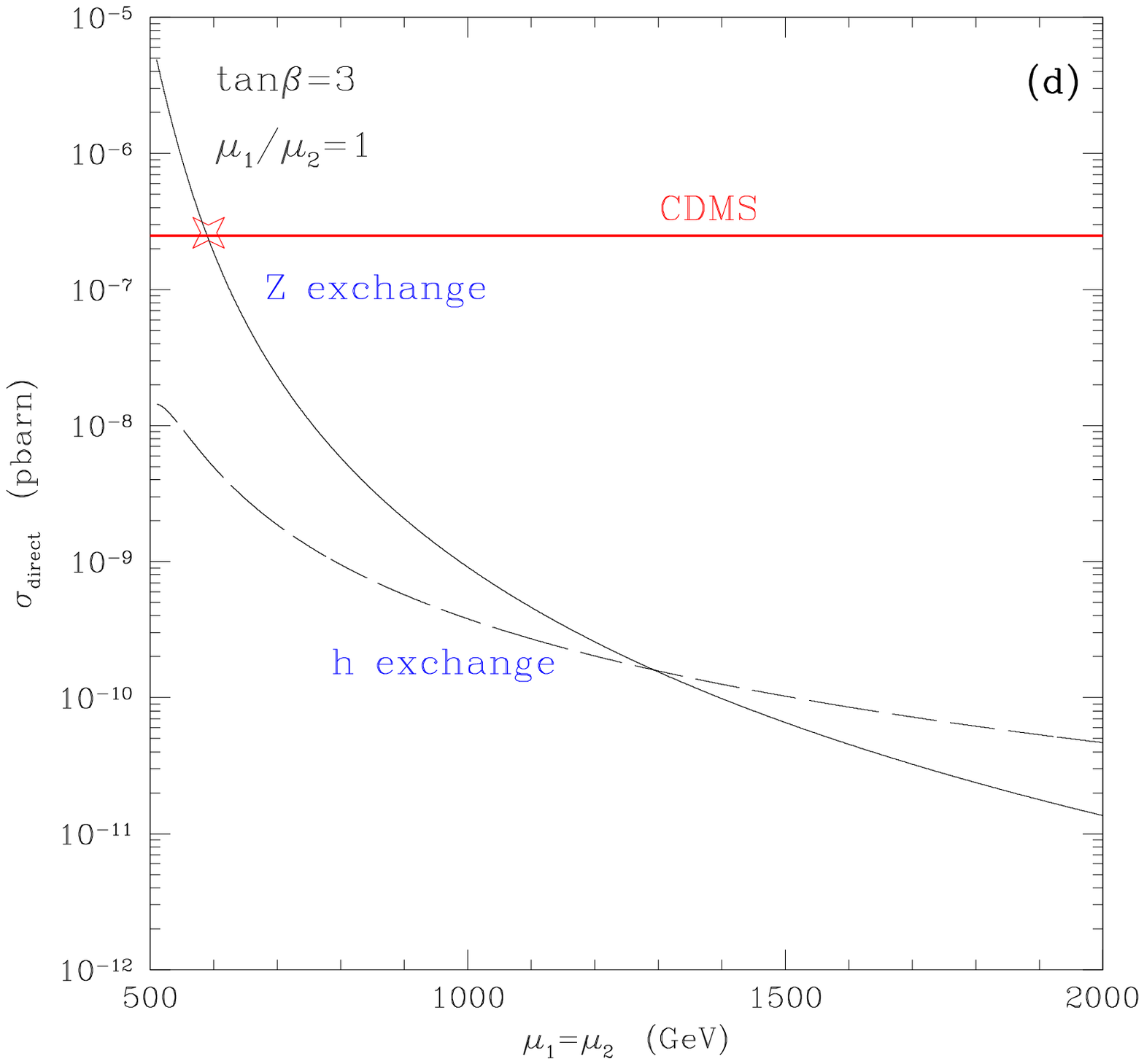}
\end{center}
\caption{Various observables calculated for the representative
  configuration $m_{\chi}=465$ GeV and $\langle\sigma
  v\rangle_{l\overline{l}}=1.4\times 10^{-24}$ cm$^3$ s$^{-1}$. (a)
  Solid line: positron fraction compared to the PAMELA
  data~\cite{PAMELA}. Dashes: secondary background (Model \#0 of
  Ref.~\cite{fermi09}); (b) Solid line: electron+positron flux
  compared to the FERMI--LAT data~\cite{fermi09}. Dashes: secondary
  background (Model \#0 of Ref.~\cite{fermi09}); (c) Solid line:
  antiproton fraction compared to the PAMELA data~\cite{pbar}. Dashes:
  secondary background~\cite{pbar_background}. The value of the
  hadronic annihilation cross section $\langle\sigma
  v\rangle_{W^+W^-}$ used for the calculation corresponds to the
  configuration indicated with a star in panel (d) which is at the
  level of the experimental CDMS II exclusion plot; (d)
  Neutralino--nucleon cross section as a function of $\mu_1=\mu_2$ for
  $\tan\beta=3$. The horizontal line is the corresponding CDMS II
  upper bound~\cite{CDMS}.}\label{fig:fluxes}
\end{figure}

The expected fluxes for positrons, electrons, and antiprotons, as
well as the expected direct detection cross section for a Dirac
neutralino are shown in detail in Fig.~\ref{fig:fluxes} for the
specific example $m_{\chi}\simeq 465$ and $\langle\sigma
v\rangle\simeq 1.4 \times 10^{-24}$ cm$^{-3}$ s$^{-1}$. This
configuration corresponds to the lower mass bound implied by the
FERMI-LAT $e^++e^-$ data if no additional sources of electrons and
positrons are assumed and, as shown in
Fig.~\ref{fig:boost_factor}, implies a boost factor of around 3 in
the case of $m_{\widetilde{l}}/m_{\chi}\simeq$1 and in the limit
of a pure bino. In particular, in Figs. \ref{fig:fluxes}a and
\ref{fig:fluxes}b, the corresponding positron and $e^++e^-$ fluxes
are plotted as a function of the energy, while the antiproton flux
and the neutralino--nucleon cross section are plotted in Figs.
\ref{fig:fluxes}c and \ref{fig:fluxes}d, respectively, for the
case of a Dirac neutralino of general gaugino--Higgsino
composition in the representative case of $\tan\beta=3$ and
$\mu_1/\mu_2$=1.
%
%
Note that in the case of a non-negligible Higgsino--gaugino mixing
the boost factor for the leptonic annihilation cross section
required to explain the PAMELA data, which is plotted in
Fig.~\ref{fig:boost_factor} in the pure gaugino limit $|\mu_i|\gg
M_1$, can be somewhat larger due to the suppression of the bino
coefficient $\mathcal{N}_{22}$. However, this effect is not
sizeable. For instance, for the maximal Higgsino composition
compatible to the CDMS II limit shown by a star in
Fig.~\ref{fig:fluxes}d the boost factor turns out to be $\simeq$4
instead of $\simeq$3.

\section{Conclusions}

Determining the Majorana/Dirac nature of gauginos will be an
interesting task for future experiments to look for supersymmetry.
Moreover, as an interesting variance of the standard Majorana
case, Dirac gauginos can be natural realizations of leptophilic
DM, which has been recently proposed to explain the rising
positron flux observed in cosmic rays by PAMELA without producing
excesses in the antiproton signal. In this paper, we have analyzed
the phenomenology of Dirac gauginos in a specific supersymmetric
realization, where a continuous R symmetry is assumed to protect
vanishingly small Majorana masses. We have shown that a light
Dirac gaugino ($m_{\chi}\lsim$ 500 GeV) can explain the PAMELA
data with moderate or no boost factor for light ($\gsim m_{\chi}$)
slepton masses. Furthermore, in the case of a non-vanishing
Higgsino fraction, Dirac gauginos can have a vector coupling with
the $Z$ gauge boson leading to a sizable spin--independent
scattering off nuclei. In some specific examples, we have shown
that present constraints on the Higgsino fraction of Dirac
gauginos from direct detection experiments are at the same level
of those coming from antiproton data. This implies that a Dirac
gaugino signal is potentially at the level of the sensitivity of
direct detection experiments at present and in the near future.

\medskip

{\bf Acknowlegement:} E.J.C. was supported by Korea Neutrino Research
Center through National Research Foundation of Korea Grant
(2009-0083526). S.S. was supported by the WCU program
(R32-2008-000-10155-0) of National Research Foundation of Korea.


\end{document}